\documentclass[aps,twocolumn,amssymb,showpacs]{revtex4}

\usepackage{graphicx}

\newcommand{\be}{\begin{equation}}
\newcommand{\ee}{\end{equation}}
\newcommand{\bea}{\begin{eqnarray}}
\newcommand{\eea}{\end{eqnarray}}

\def\p1{\pi_1}

\begin{document}

\title{Approximately self-similar critical collapse in 2+1 dimensions}
\author{Marco Cavagli\`a}
\altaffiliation{Email address: cavaglia@phy.olemiss.edu}
\affiliation{Department of Physics and Astronomy, University of Mississippi,
University, MS 38677-1848, USA}
\author{G\'erard Cl\'ement}
\altaffiliation{Email address: gclement@lapp.in2p3.fr}
\affiliation{Laboratoire de Physique Th\'eorique, LAPTH (CNRS),
B.P. 110, F-74941, Annecy-le-Vieux cedex, France}
\author{Alessandro Fabbri}
\altaffiliation{Email address: fabbria@bo.infn.it}
\affiliation{Dipartimento di Fisica dell'Universit\`a di Bologna and INFN
sezione di Bologna,
Via Irnerio 46, 40126 Bologna, Italy\\
GReCO, Institut d'Astrophysique de Paris (CNRS), 98bis Boul.\ Arago,
75014 Paris, France}

\begin{abstract}
Critical collapse of a self-gravitating scalar field in a (2+1)-dimensional
spacetime with negative cosmological constant seems to be dominated by a
continuously self-similar solution of the field equations without cosmological
constant. However, previous studies of linear perturbations in this background
were inconclusive. We extend the continuously self-similar solutions to
solutions of the field equations with negative cosmological constant, and
analyse their linear perturbations. The extended solutions are characterized by
a continuous parameter. A suitable choice of this parameter seems  to improve
the agreement with the numerical results. We also study the dynamics of the
apparent horizon in the extended background.
\end{abstract}


\maketitle

Following the numerical work of Choptuik and Pretorius on the critical collapse
of scalar matter field in 2+1 dimensions \cite{prechop} (see also
Ref.~\cite{huol}), there has been debate about understanding and analytically
reproducing their results \cite{garf,clefab,garfgund,hiwawu}. Garfinkle found a
one-parameter ($n$) family of continuously self-similar (CSS) solutions, and
proposed that one of these is the critical solution for scalar field collapse
in 2+1 dimensional AdS spacetime \cite{garf}. Numerical comparisons suggest
that the critical value is $n=4$. Subsequently, Garfinkle and Gundlach
\cite{garfgund} performed the linear perturbation analysis in this background
and found that the only solution exhibiting a single growing mode is the $n=2$
solution. Because of this discrepancy, they characterized their work as
`inconclusive'. A weak point of their approach is that it neglects the negative
cosmological constant $\Lambda$, although the latter is essential for the
existence of black hole solutions in three dimensions (the BTZ black hole
\cite{btz}). This was motivated in Ref.~\cite{garf} by the following arguments:
i) self-similarity requires $\Lambda=0$; ii) close to the singularity, the
$\Lambda$ contribution to the full solution is negligible. Although these
arguments seem reasonable, we expect the cosmological constant to play a
crucial role in black hole formation \cite{prechop}. Therefore, the inclusion
of $\Lambda$ in the above analysis may solve the above contradiction on the
critical value of $n$.\\

\noindent
The Garfinkle family of CSS solutions is
\bea
\label{garfinkle}
ds^2&=& A(v^q+u^q)^{4c^2}dudv-\frac{(v^{2q}-u^{2q})^2}{4}d\theta^2\,,
\nonumber \\
\phi &=& -2c \ln (v^q+u^q)\,,
\eea
where $A=2^{2(1-q)/q}q^2$ and $c^2=1-1/2q$.  These solutions
satisfy the three-dimensional Einstein equations
\be\label{ein}
G_{ab}-\Lambda g_{ab}=\nabla_a\phi\nabla_b\phi-\frac{1}{2}g_{ab}
(\nabla\phi)^2
\ee
with $\Lambda=0$. The source term in Eq.~(\ref{ein}) is the stress-energy
tensor of the minimally-coupled massless scalar field $\phi$. (Note that the
scalar field in Eq.~(\ref{ein}) and Ref.~\cite{garf} differ by a factor
$-1/\sqrt{4\pi}$.) The Garfinkle CSS solutions are singular at $u=v=0$. If $q$
is a positive integer $n$, the initial region  $u\ge 0$, $v\ge 0$ can be
extended across the surface $v=0$, which plays the role of an apparent horizon.
($q=n$ will be assumed below).\\

\noindent
We first extend Eq.~(\ref{garfinkle}) to solutions of the Einstein equations
with $\Lambda < 0$ and then consider the perturbation analysis in this
background. Since the cosmological constant breaks the self-similarity, the
appropriate variables are a scaling variable, for instance $u$, and
a similarity variable, which we choose as $y=(v/u)^n$. The metric
coefficients and the scalar field are expanded in terms of the dimensionless
combination $\Lambda u^{4n}$
\bea\label{extgarf}
r &\equiv&\sqrt{-g_{\theta\theta}}=r_0+\Lambda u^{6n}F(y)+\dots\,,
\nonumber\\
\sigma &\equiv& \frac{1}{2}\ln(2g_{uv})=\sigma_0+\Lambda
u^{4n} G(y)+\dots\,,
\nonumber \\
\phi &\equiv& \phi_0+\Lambda u^{4n} H(y)+\dots\,,
\eea
where $r_0$, $\sigma_0$ and $\phi_0$ are the background contributions in
Eq.~(\ref{garfinkle}). At each order in the expansion the functions $F$, $G$,
and $H$ satisfy a system of second-order coupled ordinary differential
equations. We only consider the truncation of the expansion (\ref{extgarf}) to
the first-order. The relevant equations are
\bea
& & -yF''+5F'=\frac{A}{4n^2}y^{\frac{1-n}{n}}(1-y)(1+y)^{\frac{5n-2}{n}}\,,
\label{a}\\
& & -y(1-y^2)H''+ 2(2-y^2)H'-4yH-2c\frac{1-y}{1+y}F'-
\nonumber\\
& &\qquad\qquad\qquad
-4c\frac{2+3y}{(1+y)^2}F = 0\,,
\label{b}\\
& & -yF'' + \frac{1-n +
(3n-1)y}{n(1+y)}F'-\frac{4c^2y}{(1+y)^2}F
\nonumber \\
& &\qquad\qquad\qquad -2y^2G'+2cy(1-y)H'=0\,,
\label{c}
\eea
plus two first-order contraints that reduce the moduli space of the initial
conditions. Below we briefly discuss the solutions of Eqs.~(\ref{a})-(\ref{c}).
More details will be given in Ref.~\cite{critfull}. The solution of
Eq.~(\ref{a}) regular at the center $y = 1$ ($u = v$) is
\be\label{F}
F(y)=\int_1^y y^5f(y)dy+\alpha(1-y^6)\,,
\ee
where
\be
f(y) = -\frac{A}{4n^2}\int_1^y y^{\frac{1-7n}n}(1-y)(1+y)^{\frac{5n-2}n}
dy\,.
\ee
The constant $\alpha$ can be set to zero by the gauge transformation
\be\label{gauge}
u \to u\left(1-\frac{\Lambda\alpha}{n}u^{4n}\right)\,, \quad v \to
v\left(1-\frac{\Lambda\alpha}{n}v^{4n}\right)\,.
\ee
$H$ can be obtained from Eq.~(\ref{b}). The two independent solutions of the
homogeneous equation are
\bea
H_1 &=& 3+2y^2+3y^4\,, \nonumber \\
H_2 &=& (3+2y^2+3y^4)\ln\left|\frac{1+y}{1-y}\right| -6y(1+y^2)\,.
\eea
The regular solution of the inhomogeneous equation is
\be
H=C_1H_1 + C_2H_2\,,
\ee
where
\bea
C_1 &=&\int_1^y XH_2 + c\beta\,, \ \ C_2= -\int_1^y XH_1\,,\\
X &=&\frac{c}{64y^5(1+y)^2}\left[ (1-y^2)F' + 2(2+3y)F \right]\,.
\eea
Finally, $G(y)$ is obtained by integration of Eq.~(\ref{c}) with the boundary
condition $G(1)=-F'(1)=0$. This condition follows from one of the constraints
and implies  the absence of conical singularities:
\be
g^{\mu\nu}r_{,\mu}r_{,\nu}|_{y=0}=4e^{-2\sigma}r_{,u}r_{,v}|_{y=0}=-1\,.
\ee
The functions $F$, $G$ and $H$ are shown to be analytic in $z\equiv
y^{1/n}$. \\ \\

\noindent This first-order extension of the Garfinkle solutions is
not uniquely defined, as we have found a one-parameter ($\beta$)
family of regular, analytic in $z$, solutions:
\be\label{beta}
(F,G,H)= (F, \bar G, \bar H) + c\beta (0,G_{\beta}, H_{\beta})\,,
\ee
where $\bar G$ and $\bar H$ are the $\beta=0$ solutions, and
$G_{\beta}=-c(1-y)^2(3+2y+3y^2)$, $H_{\beta}= H_1$. It can be shown
that a new integration constant will appear at each successive order
in the $\Lambda$-expansion. Thus, there is a manifold of exact
solutions of Eqs.~(\ref{a})-(\ref{c}) asymptotic to the Garfinkle
solutions near the singularity $u=0$. \\

\noindent
Let us determine the effect of the first-order $\Lambda$-corrections on the
location of the apparent horizon. From the  definition of the apparent horizon
$(\nabla r)^2=0$, we obtain, to first-order in $\Lambda$,
\be\label{apphor}
\left(\frac{y}{(1+y)^2}\right)^{2-1/n}\left(1-\Lambda
u^{4n}\psi(y)\right) = 0\,,
\ee
where
\be\label{psi}
\psi(y)\equiv 2G -6F + \frac{(1+y^2)}{y} F'\,.
\ee
For $\Lambda = 0$, the apparent horizon is the past light cone $y = 0$
of the singularity $u=v=0$. For $\Lambda < 0$, the
behavior of the functions $F$ and $G$ near $y=0$ is
\bea\label{fgappr}
F(y) &\sim & F(0) + \frac{A}{4(6n-1)} y^{1/n}\,, \nonumber \\ G(y)
&\sim & G(0) + \frac{A(8n-1)}{4(5n-1)(6n-1)}y^{1/n}\,,
\eea
where $F(0)$ and $G(0)$ are determined numerically.  On the apparent horizon,
Eqs.~(\ref{fgappr}) imply
\be\label{psi0}
u^{-4n} = \Lambda\psi(y) \simeq \frac{\Lambda A}{4n(6n-1)}y^{1/n-2}\,.
\ee
Therefore, the apparent horizon recedes into the region $z = y^{1/n}<0$ and
becomes spacelike. This feature is essentially due to the term  $F'/y$
in Eq.~(\ref{psi}) and does not depend on $\beta$. \\

\noindent
The linear perturbation analysis in this background can be performed
by expanding $r$, $\sigma$ and $\phi$ as
\bea\label{modexp}
r &=& r_0 + \Lambda u^{6n}F(y) + \epsilon u^{2n-2kn}\left[f_0(y) +
\Lambda u^{4n}f_1(y) \right]\,,
\nonumber \\
\sigma &=& \sigma_0 + \Lambda u^{ 4n}G(y) +
\epsilon u^{-2kn}\left[ g_0(y) + \Lambda u^{4n}g_1(y)\right] \,,
\nonumber \\
\phi&=& \phi_0+\Lambda u^{4n}H(y)+\epsilon u^{-2kn}\left[h_0(y)+
\Lambda u^{4n}h_1(y)\right]\,,
\eea
where $\epsilon$ is a small parameter that controls the strength of the
perturbation, and we have truncated the expansion to first-order in $\Lambda
u^{4n}$.  The growing modes are given by $Re(k)>0$. The critical solutions have
a single growing mode \cite{gundlach}. \\

\noindent
The analysis of the zeroth-order perturbations  $f_0$, $g_0$ and $h_0$ was
carried out in Ref.~\cite{garfgund}. Here, we only recall the main points of
this analysis. The regular solution of the differential equation
for $f_0$ ($f_0(1)=0$),
\be
 -yf_0'' + (1-2k)f_0' = 0 \,, \label{P01}
\ee
is
\be
f_0=c_1(1-y^{2-2k})\,.
\ee
This solution is pure gauge, i.e.\ it can be generated from the
unperturbed solution $r_0(u,v) = (u^{2n}-v^{2n})/2$ by the coordinate
transformation
\be\label{gaugtr}
u\to u\left(1+\frac{\epsilon c_1}{n}u^{-2kn}\right)\,, \ \ \
v\to v\left(1+\frac{\epsilon c_1}{n}v^{-2kn}\right)\,.
\ee
In the gauge $f_0=0$, the scalar field perturbation $u^{-2kn}h_0(y)$
solves the massless Klein-Gordon equation for the $\Lambda = 0$
background spacetime. The solution, in terms of hypergeometric
functions, depends on two integration constants. The first one is
fixed by the regularity condition, i.e.\ the absence of logarithmic
divergence for $y = 1$. The second integration constant is fixed by
the condition of smoothness on the null line $y = 0$, i.e. the
analyticity (in at least one gauge) of $h_0$ as a function of
$y^{1/n} = v/u$. A necessary condition is $2kn = m$, where $m$ is a
positive integer. For $m<n$ this condition is also sufficient. For
$n<m<2n$ one can find a gauge, i.e.\ a value of $c_1$, such that
$h_0$ is analytic. By contrast, for $m = n$ and $m \ge 2n$ there is
no gauge in which $h_0$ is analytic. The second-order equation for
$g_0$ shows that $g_0$ is generically divergent on the null line
$y=0$. However, for $1<m<n$ there is a gauge in which $g_0$ is
analytic, and for the value $m = 2n-1 > n$ ($n > 1$) $g_0$ and $h_0$
are analytic in the same gauge. Therefore, regularity at the origin
and analyticity in $y^{1/n}$ require $k=m/2n$ and either i) $1<m<n$
or ii) $m=2n-1$ ($n>1$). It can be easily seen that only the solution
with $n=2$ has a single unstable mode, namely $m=3$ ($k=3/4$). \\

\noindent The only debatable question in this analysis is whether the
requirement that the non-scalar quantity $g_0$ is analytic at $y = 0$
might not be too strong. In principle, it should be enough to demand
that the perturbation of a scalar quantity, such as the Ricci scalar,
is analytical at $y = 0$. At zeroth-order in $\Lambda$, the Ricci
scalar is
\be
R = R_0(u,y)(1-2\epsilon u^{-2kn}\rho_0(y))\,,
\ee
where \cite{critfull}
\be
\rho_0 = g_0 + \frac{1+y}{4c}[(1-y)h_0' - 2kh_0]\,.
\ee
In the gauge $c_1 = 0$, $g_0$ and $h_0'$ diverge for $y \to 0$ as
$y^{-m/n}$. Therefore, $\rho_0$ diverges. However, the zeroth-order
Ricci scalar $R_0$ goes to zero as $y^{1-1/n}$ and the perturbation
\be
\delta R \propto R_0\rho_0 \sim u^{-4n-m}y^{1-(m+1)/n}\,,
\ee
remains finite at $y=0$ for $m < n$, including $m = 1$. If the mode
$m = 1$ ($k = 1/2n$) were allowed, none of the Garfinkle solutions
would be critical: for $n=1$ there would be no growing mode, for
$n=2$ there would be two growing modes, $m=1$ and $m=3$, etc.
However, as we now show, the extra modes with $m=1$ do not survive
the first-order extension in $\Lambda$.
\\

\noindent
The first-order perturbation $f_1$ solves the inhomogeneous
differential equation
\be
-yf_1'' + (5-2k)f_1' =
\frac{A}{2n^2}y^{\frac{1-n}{n}}(1+y)^{\frac{2(2n-1)}n}[f_0+(1-y^2)g_0]\,.
\ee
In the gauge $f_0=0$, $g_0 \sim y^{-m/n}$ implies
\be
f_1 \sim y^{\frac{1-m}n}\quad  (m>1) \quad \mbox{\rm or} \quad \ln y
\quad (m=1)
\ee
for $y\to0$. If $m>1$, the divergence of $f_1$ can be gauged away by the
zeroth-order gauge transformation (\ref{gaugtr}). The logarithmic divergence of
the first-order contribution to the $m=1$ mode cannot be gauged away; this mode
is never analytic at $y=0$. A detailed analysis of the first-order
perturbations will be presented elsewhere \cite{critfull}. Here, let us just
note that the analytic and numerical integrations of the first-order
perturbations indicate that all the modes found in the analysis of the
zeroth-order perturbations satisfy the boundary conditions of regularity and
analyticity at first-order for any value of $\beta$. This shows that the
analysis of Garfinkle and Gundlach \cite{garfgund} is robust, i.e.\ it survives
extension to first-order in $\Lambda$.\\

\noindent
Now let us discuss the effect of the extension on the behavior of the apparent
horizon for the perturbed critical solution ($n=2$, $k=3/4$). The
apparent horizon satisfies the equation
\be\label{apphor1}
\frac{y^{3/2}}{(1+y^3)}\bigg(1-\Lambda u^{8}\psi -\epsilon u^{-3}\chi
-\epsilon\Lambda u^{5}\eta\bigg)=0 \,,
\ee
where $\psi$ is defined in Eq.~(\ref{psi}) and
\be\label{chi}
\chi = 2g_0 -\frac12f_0 +\frac{1+y^2}{y}f_0'\,.
\ee
(The exact form of $\eta$ is inessential for the following discussion and will
be given in  Ref.~\cite{critfull}.)  For $y\to 0$, $\chi$ is dominated by the
last term in Eq.~(\ref{chi}):
\be\label{chi0}
\chi \simeq -\frac{c_1}2\,y^{-3/2}\,.
\ee
The zeroth-order approximation of Eq.~(\ref{apphor1}) with $\Lambda = 0$ is
\be
v^3 \simeq -\frac{\epsilon c_1}2\,.
\ee
The apparent horizon is null, and exists for both signs of
$\epsilon$. (The singularity $u=v=0$ is hidden by the apparent
horizon only for $\epsilon c_1< 0$.) The situation changes dramatically
when we take into account the first-order contributions. Neglecting
the term $\eta$, we see from Eq.~(\ref{psi0}) and Eq.~(\ref{chi0}) that near
$y=0$ the shape of the apparent horizon is determined by a balance
between the $\Lambda$ and $\epsilon$ contributions. The leading behavior is
\be
 u \simeq u_0 \left(1-\frac{y}{3}\right)\,, \quad u_0\equiv
\left(\frac{22\epsilon c_1}{\Lambda}\right)^{1/11}\,.
\ee
The apparent horizon, which exists only for $\epsilon c_1<0$, is spacelike for
small positive $y$ and becomes null ($u=u_0$) for $y=0$. The numerical solution
of Eq.~(\ref{apphor1}) shows that on the apparent horizon $u$ is everywhere
bounded by $u_0$. This confirms a posteriori that the $\eta$ contribution to
Eq.~(\ref{apphor1}) can be neglected for small $\epsilon$. The existence and
the shape of the apparent horizon, which hides the singularity $u=v=0$, do not
depend on the parameter $\beta$. \\

\begin{figure}
\includegraphics[angle=270,width=3.4in,clip]{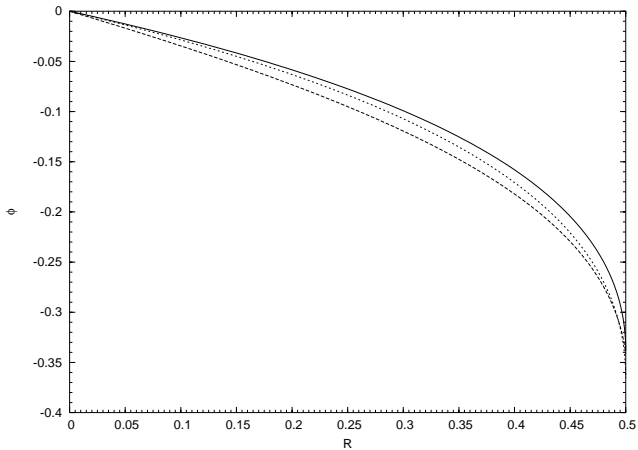}
\caption{The scalar field $\phi$ as a function of $R$ for the $n=2$ CSS
solution (solid), the $n=2$ extended CSS solution (dashed), and the $n=4$ CSS
solution (dotted). The $n=2$ extended CSS solution is computed for $\beta=1$
and $T_0=1.9$. The scalar field has been rescaled by a factor $-1/\sqrt{4\pi}$
to facilitate comparison with the results of Ref.~\cite{garf}.}
\label{fig1}
\end{figure}

\noindent
Finally, we present some evidence that the $O(\Lambda)$ corrections
improve the agreement with the numerical simulations of near-critical collapse.
Following Ref.~\cite{garf}, we introduce the coordinates $(T,R)$
\be
T = -2n\ln u \,, \ \ \
R=u^{-2n}r=\frac{1-y^2}{2} +\Lambda e^{-2T}F(y)\,.
\ee
The expression of the extended Garfinkle scalar field at some fixed
$T_0$ is
\be\label{phiext}
\phi_n(y,T_0)=-2c\ln\frac{(1+y)}{2}+\Lambda e^{-2T_0}(H(y)-H(1))\,,
\ee
where $\phi$ has been shifted by a constant to make it vanish at $y=1$. In
Ref.~\cite{garf} Garfinkle shows that the nonextended solution with $n = 4$
agrees with the numerical critical solution of Ref.~\cite{prechop} at an
intermediate time $T_0\sim 9$. For such a large $T_0$, the extended $\phi_n$
(\ref{phiext}) reduces to that of the CSS solution. However, the calibration of
the numerical $T_0$ involves some ambiguity. In Ref.~\cite{prechop}, $T$ is
defined by $T = -\ln t_c$, where $t_c = 0$ ($T = +\infty$) at the accumulation
point (the singularity). Even a tiny error in the determination of this zero
from near-critical simulations will translate into a large error on the
corresponding value of $T_0$. Therefore, the latter has to be considered as an
unknown parameter. A second unknown parameter in Eq.~(\ref{phiext}) is
$\beta$.  $T_0$ and $\beta$ can be set by comparing $\phi_2(0,T_0)$ for the
critical solution $n=2$ with $\phi_4(0,\infty)$. From the numerical solution of
Eqs.~(\ref{a})-(\ref{c}), we find $\bar{H}(0) \sim 0.016 \ll c\beta
H_{\beta}(0)=3c\beta$, provided that $\beta$ is not too small. So
$\phi_2(0,T_0)$ depends only on the product $\beta e^{-2T_0}$. By comparing the
latter with the numerical value of $\phi_4(0,\infty)$, we obtain $\beta
e^{-2T_0}\sim 0.022$. For the (arbitrary) choice $\beta=1$ this gives $T_0\sim
1.9$. In Fig.\ 1 we plot in terms of $R$ the CSS solution with $n=2$, the
extended solution for $n=2$ and the CSS solution with $n=4$. The $O(\Lambda)$
corrections seem to improve the agreement of the $n=2$ critical solution with the
numerical results.
\begin{figure}
\includegraphics[angle=270,width=3.4in,clip]{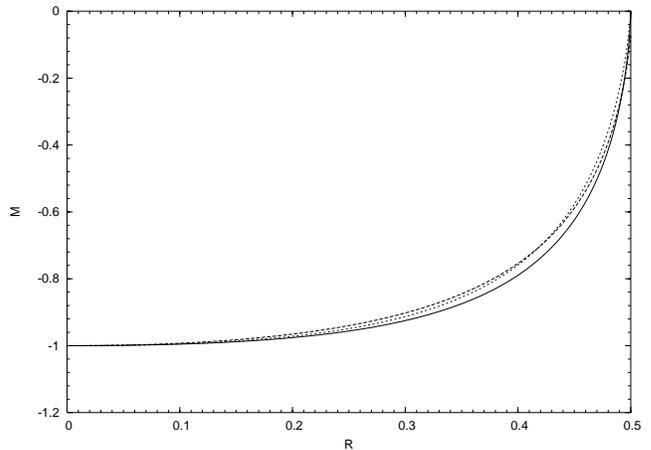}
\caption{The mass aspect $M$ as a function of $R$ for the $n=2$ CSS
solution (solid), the $n=2$ extended CSS solution (dashed), and the $n=4$
CSS solution (dotted).}
\label{fig2}
\end{figure}
This conclusion is strengthened by Fig.\ 2, where we plot
the mass aspect in terms of $R$
\bea\label{mass}
M_n(y,T_0) &\equiv & -\Lambda r^2+4e^{-2\sigma}r_{,u}r_{,v}= -\left[
\frac{4y}{(1+y)^2}\right]^{\frac{2n-1}n}
\nonumber \\
&&\null\hskip -0.6in -\Lambda u^{4n} \left[ \frac{(1-y^2)^2}{4}-
\left(\frac{4y}{(1+y)^2}\right)^{\frac{2n-1}{n}}\psi(y)\right]\,,
\eea
for the same values of $\beta$ and $T_0$. However, it is clear that the
first-order extended $n=2$ solution agrees with the numerics only over a small
range of $T_0$, as opposed to the $n=4$ CSS solutions, which agree over a large
range of intermediate times \cite{garf}. This suggests that the question of the
agreement between the analytical and numerical critical solutions is still an
open problem. \\

\noindent
To conclude, our analysis shows that in the near-critical regime the shape of
the apparent horizon is determined by a balance betwen the $\Lambda$ and
$\epsilon$ contributions. This is evidence that the cosmological constant plays
a role in black hole formation. We have also shown that the apparent
contradiction between the results of Ref.~\cite{prechop} and Ref.~\cite{garf}
can partly be solved by including $O(\Lambda)$ terms. Another result of our
analysis is that, at this order, there seems to be a one-parameter family of
critical solutions, rather than a single critical solution. This parameter is
not connected with gauge transformations (the gauge parameter is $\alpha$,
which has been set to zero). Rather, as will be discussed in more detail in
Ref.~\cite{critfull}, the first-order terms linear in $\beta$ in
Eq.~(\ref{beta}), which solve the  homogeneous equations (\ref{a})--(\ref{c}),
can be reinterpreted as zeroth-order $k=-2$ perturbations. It follows that the
extended $n=2$ critical solution is unique modulo the addition of a decaying 
perturbation. 



\begin{thebibliography}{99}

\bibitem{prechop}
F.\ Pretorius and M.W.\ Choptuik, {\it Phys.\ Rev.} D {\bf 62}, 124012 (2000)
\bibitem{huol}
V.\ Husain and M.\ Olivier, {\it Class.\ Quant.\ Grav.} {\bf 18}, L1
(2001)
\bibitem{garf}
D.\ Garfinkle, {\it Phys.\ Rev.} D {\bf 63}, 044007 (2001)
\bibitem{garfgund}
D.\ Garfinkle and C.\ Gundlach, {\it Phys.\ Rev.} D {\bf 66}, 044015 (2002)
\bibitem{clefab}
G.\ Cl\'ement and A.\ Fabbri,  {\it Class.\ Quant.\ Grav.} {\bf 18}, 3665 (2001);
{\it Nucl. Phys.} {\bf B630}, 269 (2002)
\bibitem{hiwawu}
E.W.\ Hirschmann, A.\ Wang and Y.\ Wu, {\it Class.\ Quant.\ Grav.} {\bf
  21}, 1791 (2004)
\bibitem{btz}
M.\ Ba\~{n}ados, C.\ Teitelboim and J.\ Zanelli, {\it
  Phys.\ Rev.\ Lett.} {\bf 69}, 1849 (1992)
\bibitem{critfull} M.\ Cavagli\`a, G.\ Cl\'ement and A.\ Fabbri, in preparation
\bibitem{gundlach} C.\ Gundlach, {\it Phys.\ Rept.} {\bf 376}, 339
  (2003)
\end{thebibliography}
\end{document}